\documentclass[12pt,preprint]{aastex}

\usepackage{natbib}
\bibpunct{(}{)}{;}{a}{}{,}
\usepackage{lscape}

\shorttitle{A search for dust emission in the Leo intergalactic cloud}
\shortauthors{Bot C. et al.}
\date{\today}

\begin{document}


\title{A Search for Dust Emission in the Leo Intergalactic Cloud\altaffilmark{*}}

\altaffiltext{*}{This work is based on observations made with the \emph{Spitzer Space Telescope}, which is operated by the Jet Propulsion Laboratory, California Institute of Technology, under a contract with NASA.}


\author{Caroline Bot\altaffilmark{1,2} and George Helou\altaffilmark{1}  and William B. Latter\altaffilmark{3}  and J\'er\'emie Puget\altaffilmark{1} and Stephen Schneider\altaffilmark{4} and Yervant Terzian\altaffilmark{5}}
\email{bot@astro.u-strasbg.fr}


\altaffiltext{1}{California Institute of Technology, Pasadena CA 91125, USA}
\altaffiltext{2}{UMR7550, Centre de donn\'ees Astronomiques de Strasbourg (CDS), 67000 Strasbourg, France}
\altaffiltext{3}{NASA Herschel Science Center, California Institute of Technology, Pasadena CA 91125, USA}
\altaffiltext{4}{Department of Astronomy, Universtiy of Massachusetts, Amherst, MA 01003, USA}
\altaffiltext{5}{Department of Astronomy/NAIC, Cornell University, Ithaca, NY 14853, USA}


\begin{abstract}
We present a search for infrared dust emission associated with the Leo cloud, a large intergalactic cloud in the M96 group. Mid-infrared and far-infrared images were obtained with IRAC and MIPS on the Spitzer Space Telescope. Our analysis of these maps is done at each wavelength relative to the H{\sc i} spatial distribution. We observe a probable detection at 8 $\mu$m and a marginal detection at 24$\mu$m associated with the highest H{\sc i} column densities in the cloud. At 70 and 160$\mu$m, upper limits on the dust emission are deduced. The level of the detection is low so that the possibility of a fortuitous cirrus clump or of an overdensity of extragalactic sources along the line of sight can not be excluded. If this detection is confirmed, the quantities of dust inferred imply a dust to gas ratio in the intergalactic cloud up to a few times solar but no less than 1/20 solar. A confirmed detection would therefore exclude the possibility that the intergalactic cloud has a primordial origin. Instead, this large intergalactic cloud could therefore have been formed through interactions between galaxies in the group. 

\end{abstract}


\keywords{ISM: clouds -- ISM: individual (Leo cloud) -- galaxies: intergalactic medium -- infrared: ISM}



\section{Introduction}

The large intergalactic cloud in the Leo group was first discovered serendipitously in H{\sc i} emission with Arecibo \citep{Schneider:1983yw}. It lies between the galaxies M105 and M96 but is clearly distinct from any galaxy of the Leo group. Further observations showed smaller isolated H{\sc i} clouds that revealed a 200kpc diameter ring shape, with radial velocities consistent with a Keplerian orbit around M105 and NGC3384 \citep{Schneider:1985pp}. The total mass of the intergalactic cloud as determined from H{\sc i} data is $\sim 10^9$--$10^{10}M_\odot$, i.e significant with respect to the H{\sc i} in all the galaxies of the group. Higher resolution observations were obtained with the Very Large Array \citep{Schneider:1986yy} which showed a clumpy structure in the main body of the cloud. The clumps reach column densities of about $4\times 10^{20}$cm$^{-2}$, central volume densities of $0.1$cm$^{-3}$ and appear to be distinct virialized entities within the cloud complex. These characteristics make them potential sites for star formation but until recently no stellar component had been observed to be associated with these fairly dense clumps \citep{Schneider:1989rt}. Although a tentative detection of H$\alpha$ was reported in the Leo intergalactic cloud  by \citet{Reynolds:1986db}, this was not confirmed by further studies in H$\alpha$ \citep{Donahue:1995vn} which found a very low upper limit for the H$\alpha$ surface brightness ($1.6\times 10^{-19} \mathrm{erg.s}^{-1}\mathrm{cm}^{-2}.\mathrm{arcsec}^{-2}$). However, a very recent study by \citet{Thilker:2009kx} reports tantalizing evidence for UV emission from parts of the Leo Cloud. This UV detection is attributed to recent massive star formation in the Leo Cloud.

Two distinct possibilities were initially discussed for the origin of the intergalactic cloud: it could be the product of head-on collision between two galaxies \citep{Rood:1985kx} or be a primordial remnant of the group's formation. The former possibility is mainly supported by evidence of tidal interactions in NGC3384 and M96 and an extension in the H{\sc i} cloud pointing at M96\citep{Schneider:1985pp}. However, the substantial mass of the cloud with respect to that of the surrounding galaxies, the lack of detected star formation in the cloud, and the minimum age of the cloud deduced from the large orbital period of the ring argued for a primordial origin. This age estimate might be questioned since it could be difficult for it to remain stable against tidal disruption by encounters, given that the crossing time of the galaxies in the group is less than the estimated age of the cloud. But the kinematics of the intergalactic ring and those of the surrounding galaxies are consistent with ordered rotation \citep{Schneider:1985pp,Schneider:1989zn} which makes collisions less likely and argues for a common origin for the galaxies of the group and the ring \citep{Schneider:1989rt}. A galaxy in the group might have undergone repeated tidal encounters during its orbit, the fragments of which have been distributed along its orbital path. \citet{Bekki:2005eu} model such a scenario with a hydrodynamical simulation. They show that the gas from the outer part of a low-surface-brightness galaxy may be stripped within the potential of a galaxy group, leading to the formation of a large gas ring with no stars. Their simulation suggests that gas removed this way may settle into a feature similar to the Leo ring after about 6~Gyr. Alternatively, in the light of their new UV detection, \citet{Thilker:2009kx} propose that the complexes could be dwarf galaxies observed during their formation. The origin of the Leo intergalactic cloud is therefore still not settled. Detecting dust emission in the Leo Cloud could then determine whether this gas is of primordial origin or not. 

This paper presents observations with the Spitzer Space Telescope taken in the direction of the Leo intergalactic cloud. After a discussion of the data reduction, the infrared emission observed in this region is compared to the H{\sc i} column densities and the origin of the emission is discussed.

\section{Observations}

\subsection{Leo Cloud IRAC and MIPS observations}

\begin{figure}
\plotone{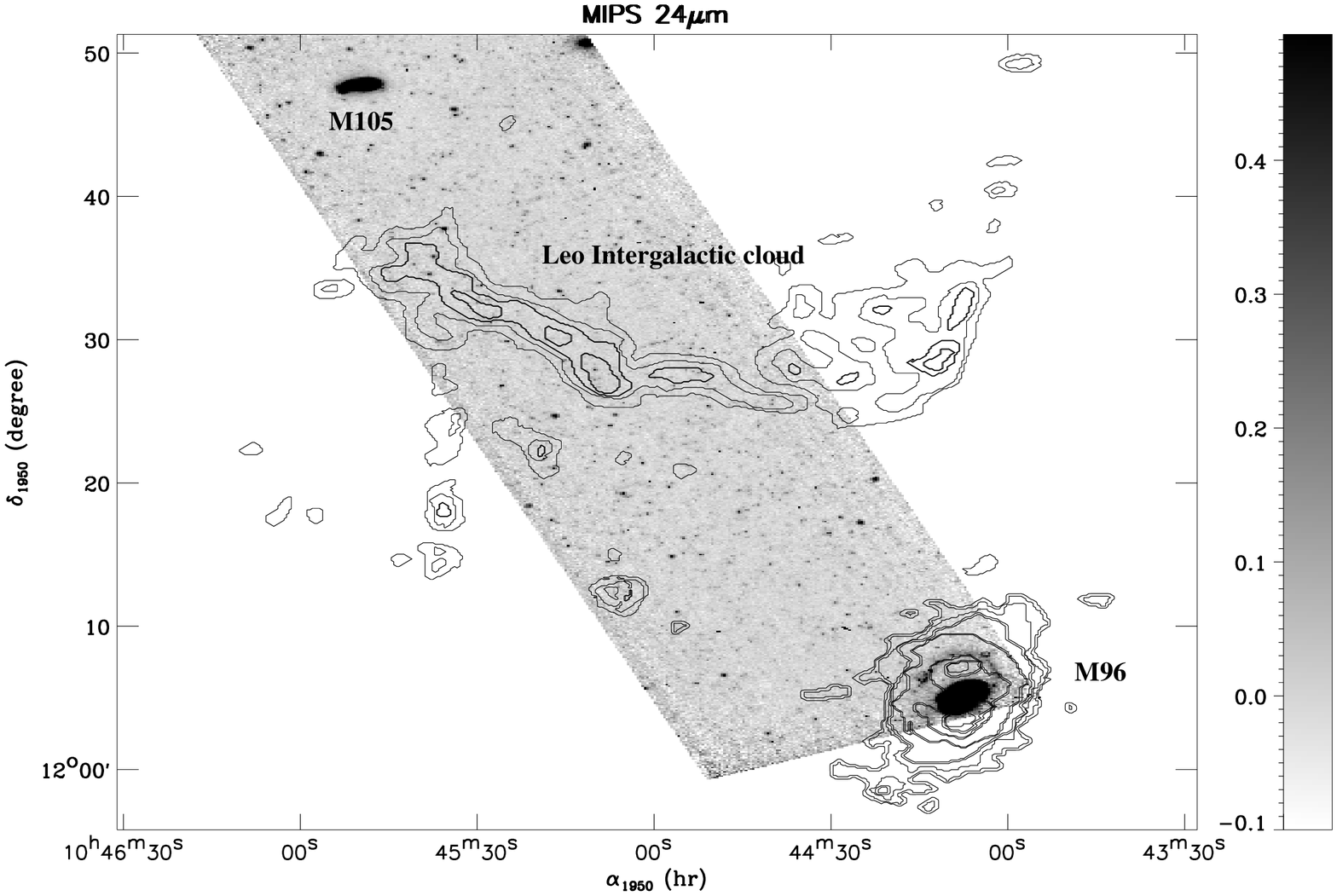}
\caption{Presentation of the regions observed: the 24$\mu$m map is shown with H{\sc i} contours overlaid. The Leo intergalactic cloud as well as the two bright galaxies surrounding it are labeled.\label{fig1}}
\end{figure}

IRAC \citep{Fazio:2004fj} and MIPS \citep{Rieke:2004zr} imaging of the Leo intergalactic cloud were obtained by the Spitzer Space Telescope program P20626 ("Search for Infrared emission from the the Leo Extragalactic H{\sc i} Cloud", P.I. George Helou). The IRAC observations targeted two of the highest column density clumps \citep[as observed in H{\sc i};][]{Schneider:1989zn}, while the MIPS observations covered a larger region of emission around these clumps (c.f Fig \ref{fig1}).

Each of the two IRAC fields were observed twice and a standard reduction of the data was applied starting at the Basic Calibrated Data (BCD) level obtained from the Spitzer data archive: data artifacts like muxbleed and column pulldown were removed, as well as outliers observed in the overlapping BCD images. In addition, we removed a linear plane in the image fitted on each BCD with point sources masked. The individual images in each observation and the two observations in each field were then combined into a mosaic. 

The MIPS data was also taken at the BCD level and combined into a mosaic for each AOR (Astronomical Observation Request). For the MIPS 24 and 70$\mu$m maps, the first frame observations were not included in the mosaics. At all wavelengths, an illlumination correction was performed in each Data Collection Event (DCE). For each AOR, a mosaic was created and a linear plane was fitted to the mosaic (with point sources masked) and removed from the data. For the 70 and 160 $\mu$m individual maps (one for each AOR), a column average was removed to reduce the impact of long term latents. This process will tend to remove extended structure along the scan length, but should not affect our study since the densest Leo Cloud structures are almost perpendicular to it. Outliers between AORs were flagged and removed, before combining all AORs into a large mosaic. For the 24$\mu$m map, a linear plane is then removed from the total mosaic (the fit is done with the sources masked) in order to be consistent to the other wavelength treatment.

\subsection{H{\sc i} data}

We compare the infrared data to H{\sc i} data from the VLA, which had a resolution of approximately 45". In the region that we sample with the infrared observations, 70 to 100 \% of the emission observed on large scales with Arecibo was recovered by these VLA observations \citep{Schneider:1986gm}, so the VLA signals are affected only slightly by missing flux issues often associated with interferometry. Since the original data are no longer accessible in digital form, the maps from the figures in \citet{Schneider:1986gm} and \citet{Schneider:1989rt} were used to create the astrometry from the coordinates given in the figure and the H{\sc i} emission from the contours. Consequently, we can only compare the infrared data with H{\sc i} in bins of column density. We assign to each bin the value corresponding to the lower contour, which means that the H{\sc i} column densities quoted for this study are underestimated in each bin.

\subsection{Point source removal and convolution}

This study aims at detecting very faint and diffuse emission. Removing point sources and any emission associated with them is therefore critical.

\begin{figure}
\epsscale{.80}
\includegraphics[scale=0.5,angle=90]{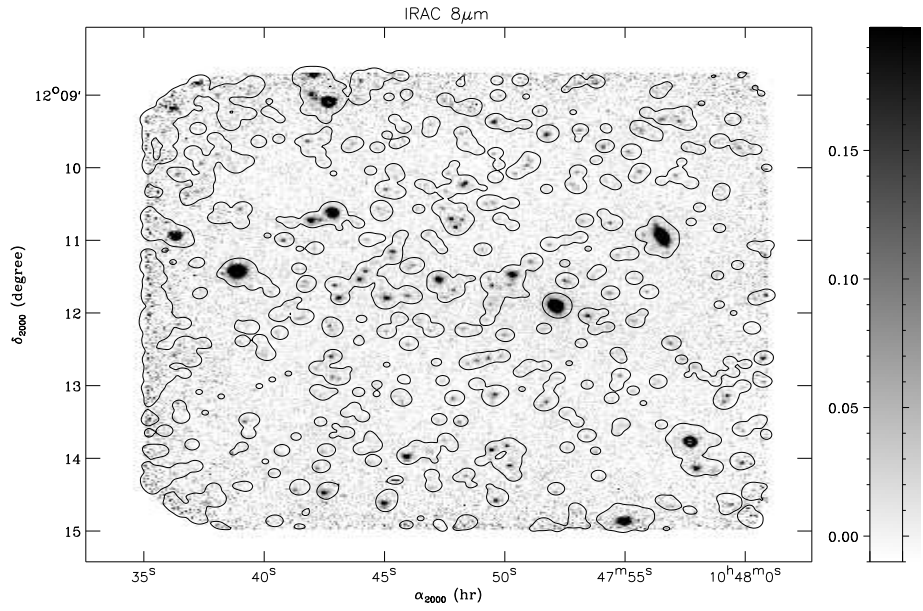}
\caption{Example of an IRAC field, at the original resolution. The contour corresponds to the mask that we  applied to exclude the point sources before the convolution of the image to the H{\sc i} resolution.\label{fig2}}
\end{figure}

This problem is most pronounced at 8$\mu$m, where foreground stars from our galaxy and distant galaxies create most of the emission observed in each field. Point sources were identified using two complementary schemes: a sigma threshold clipping, as well as a comparison with the IRAC 4.5$\mu$m map covering the same field (we create a point source mask under the assumption that all the sources detected at 4.5$\mu$m come from stars or distant galaxies). The resulting masks are convolved with a gaussian kernel (with a FWHM twice the resolution) to efficiently mask possible artifacts or large PSF wings around detected point sources. An example of the resulting point source mask at 8$\mu$m is shown for one field in Fig. \ref{fig2}.

At 24, 70 and 160$\mu$m, the number of resolved point sources is reduced, but data artifacts like stripes complicate their detection. We used the DAOPHOT-based algorithm, find, in IDL to detect most of the point sources. The galaxies with large angular sizes (M105, M96) were removed by hand, as well as a few galaxies not detected by the algorithm (e.g. embedded in a stripe). Here again, the point source mask created at each wavelength is convolved by a kernel to include extended emission around the point sources.

The point sources removed are most probably a mix of foreground stars and background galaxies in the field, but this process could remove star forming regions associated to the Leo Cloud. However, we do not observe any spatial correlation between the point source distribution and the H{\sc i}. Furthermore, H$\alpha$ observations in the direction of the cloud gave a very low upper limit on the emission \citep{Donahue:1995vn} and the star forming regions reported in the UV \citep{Thilker:2009kx} are not seen at 8$\mu$m.

\begin{figure*}
\plotone{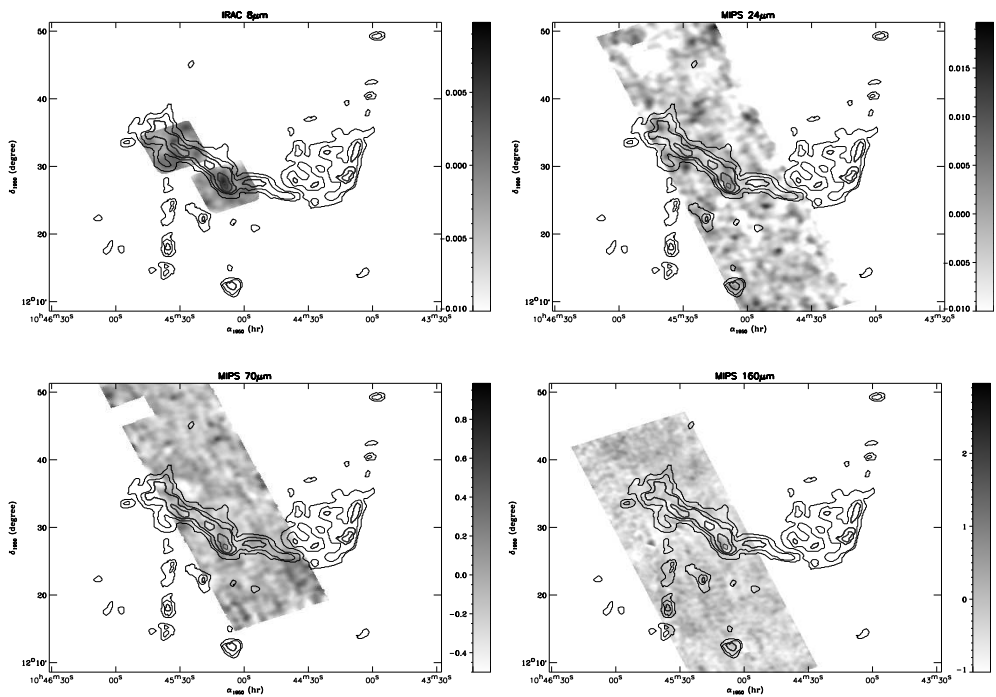}
\caption{Infrared maps at 8, 24, 70 and 160$\mu$m in the Leo Cloud region with the sources removed and convolved to the H{\sc i} resolution.  The H{\sc i} contours at 0.1, 0.5, 1.1, 2.2 and $3.8\times 10^{20}$ at cm$^{-2}$ are overlaid. \label{fig3}}
\end{figure*}

The different images with the point sources masked are then convolved to the H{\sc i} resolution. For this, we assume that the point spread function of the instruments are gaussian with full width half maximums of 1.9", 6", 17" and 37" at 8, 24, 70 and 160 $\mu$m respectively and of 45" for the H{\sc i} observations. The convolution kernel used in each case was normalized to take into account the masked region at each wavelength. The maps obtained at each wavelength are presented in Fig. \ref{fig3} with the H{\sc i} contours overlaid.

\section{Analysis of the infrared emission}

The dust emission observed at all wavelengths is very low: the variations in the infrared emission observed at each wavelength have standard deviations of 0.002, 0.008, 0.12 and 0.34MJy/sr at 8, 24, 70 and 160$\mu$m respectively. This emission originates from several potential components: 
\begin{itemize}
\item dust emission from the intergalactic cloud, 
\item residuals from data artifacts: this is particularly salient at 70 and 160$\mu$m where long term transients create stripes in the data that are difficult to remove.
\item variations in the cosmic infrared background emission, CIB: at small scales like the one sampled by Spitzer, the distant unresolved galaxies are not uniformly distributed. They can be viewed as the sum of a Poisson distribution and a clustered part following the large scale structures of the Universe.
\item foreground cirrus emission from our own galaxy: the Leo intergalactic cloud is at high galactic latitude -- +57$^o$ -- and cirrus emission is therefore expected to be faint. Furthermore, most of the emission on large scale will be erased by the plane fittings in our reduction scheme. However, small scale variations of infrared cirrus emission could exist \citep{Bot:2009sf} and create variations of the dust emission in the Leo Cloud direction that are not associated with the Leo cloud itself. Unfortunately, there is no map of the Milky Way H{\sc i} emission at sufficiently high resolution in the field of view to constrain further these variations.
\end{itemize}
The challenge is to separate all these mixed components in order to extract information on the dust emission associated with the Leo intergalactic cloud. 

\subsection{Infrared emission associated with the H{\sc i} cloud}

In this context, the information contained in the H{\sc i} data is crucial to search for emission associated with the cloud. Indeed, if dust is present in the intergalactic cloud, then it is expected to follow the same spatial distribution as the gas and the intensity of the dust emission would scale with the gas column density.

\begin{figure*}
\plottwo{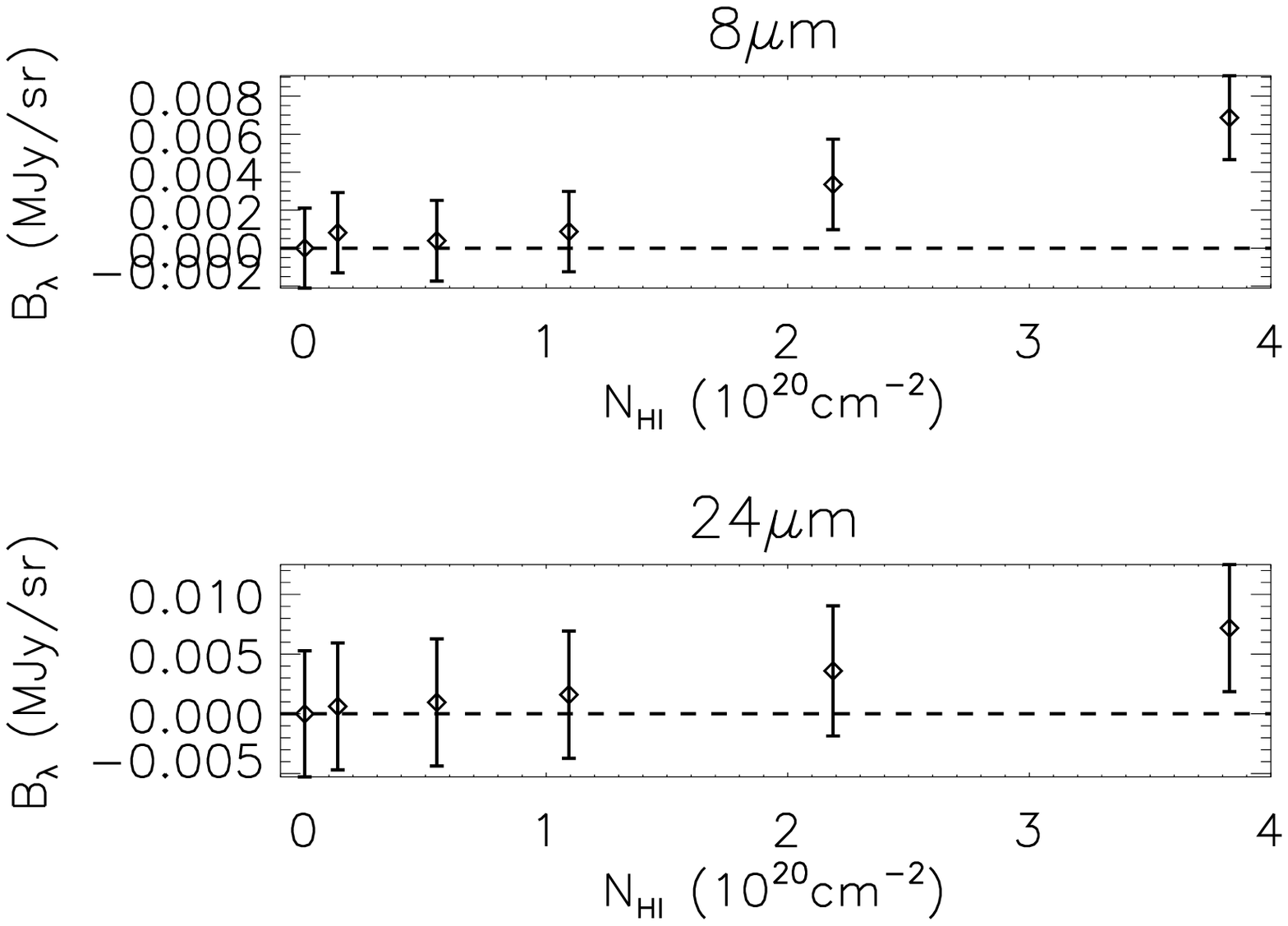}{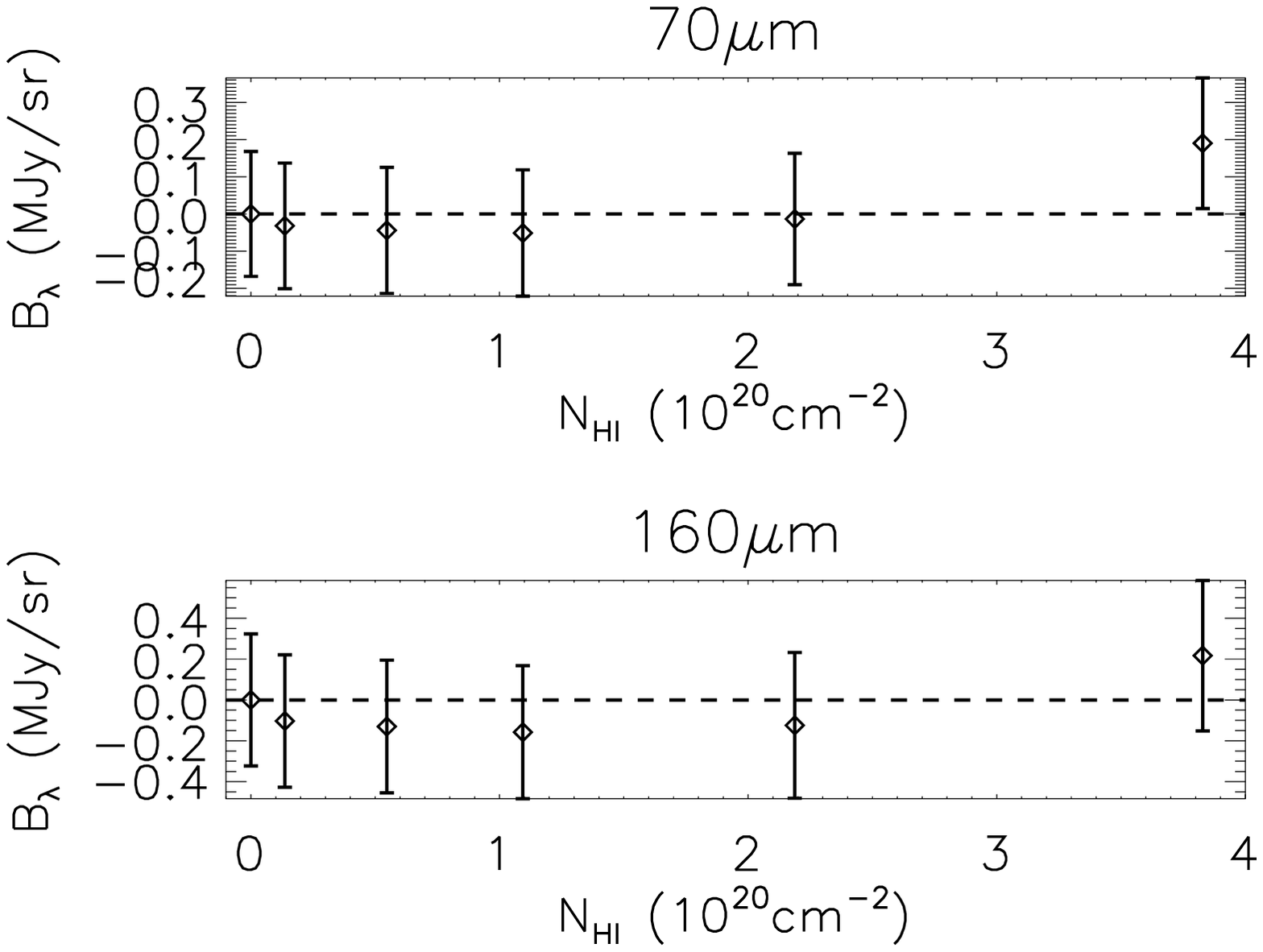}
\caption{Average dust emission (and the associated error) in the background region and in 5 H{\sc i} bins in the region of the Leo Intergalactic Cloud \label{fig4}}
\end{figure*}

Due to the instrument configurations, the area mapped at each wavelength is different. In particular, the 8$\mu$m map only covers the densest clumps of the Leo Cloud. To make the most out of the observations, each wavelength is analyzed separately, but the same scheme is applied: 
\begin{itemize}
\item the background region is defined as being the region where emission has been observed but no H{\sc i} has been detected at the Leo Cloud velocities, and the average emission observed in this background region is removed from the whole map.
\item In each H{\sc i} column density bin, the average brightness observed is computed. The error associated with the average in each bin is defined as the quadratic sum of the error on the mean (because the maps are oversampled, it is computed only on independent pixels) and of the standard deviation of the emission in the background region (hereafter, we refer to this quantity as "background variations"). Such a definition of the error therefore includes our uncertainty on the variations in the brightnesses due to foreground cirrus emission, the variations in the CIB, etc.
\end{itemize}
Fig. \ref{fig4} shows the mean infrared brightness at each wavelength as a function of the observed H{\sc i} column density from the Leo Cloud.

We observe the detection of an excess of emission in the highest H{\sc i} column density bin at 8$\mu$m ($3\sigma$ detection), a possible detection at 24$\mu$m ($1\sigma$ detection) and no clear detection above the background variations at 70 and 160$\mu$m. We estimate upper limits corresponding to 3 times the background variations on these maps. 

\subsection{Check for a spurious detection}

\begin{figure}
\epsscale{.80}
\includegraphics[scale=.5,angle=90]{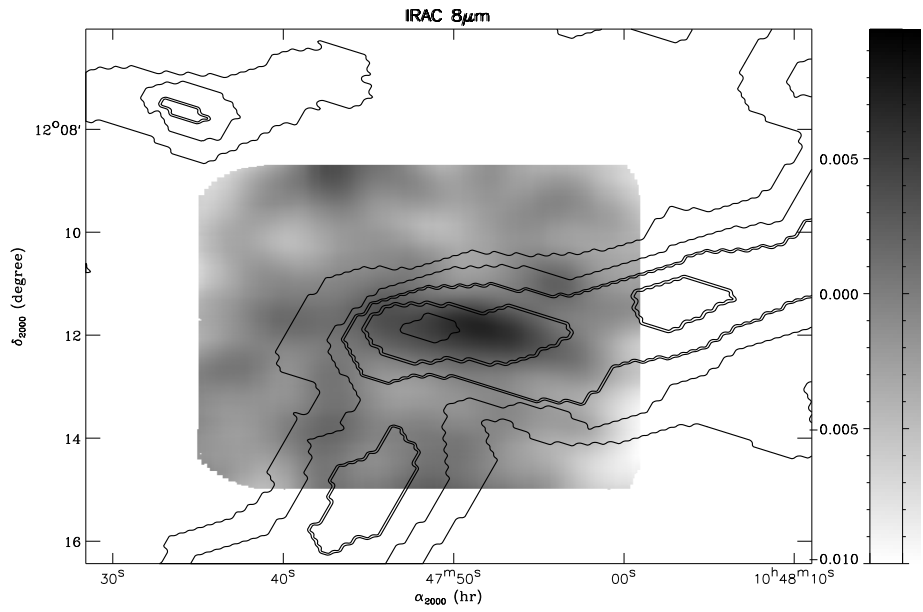}
\caption{Close-up on the 8$\mu$m IRAC image where emission associated with the H{\sc i} emission is detected. The H{\sc i} contours are overlaid with levels as in Fig. \ref{fig3}. The infrared emission shows a similar spatial distribution to the highest H{\sc i} contours, but the peaks in emission are slightly offset.   \label{fig7}}
\end{figure}

The detection at 8$\mu$m (and even more at 24$\mu$m) is barely above the background variations and appears only in the highest H{\sc i} density bin. This column density bin corresponds to only one clump of the intergalactic cloud (the densest) and a close inspection of the 8$\mu$m spatial distribution with respect to the one observed with H{\sc i} reveals that the infrared emission peak and the one seen in the H{\sc i} map are slightly shifted (c.f. Fig. \ref{fig7}). 

\begin{figure}
\epsscale{.80}
\includegraphics[scale=.5,angle=90]{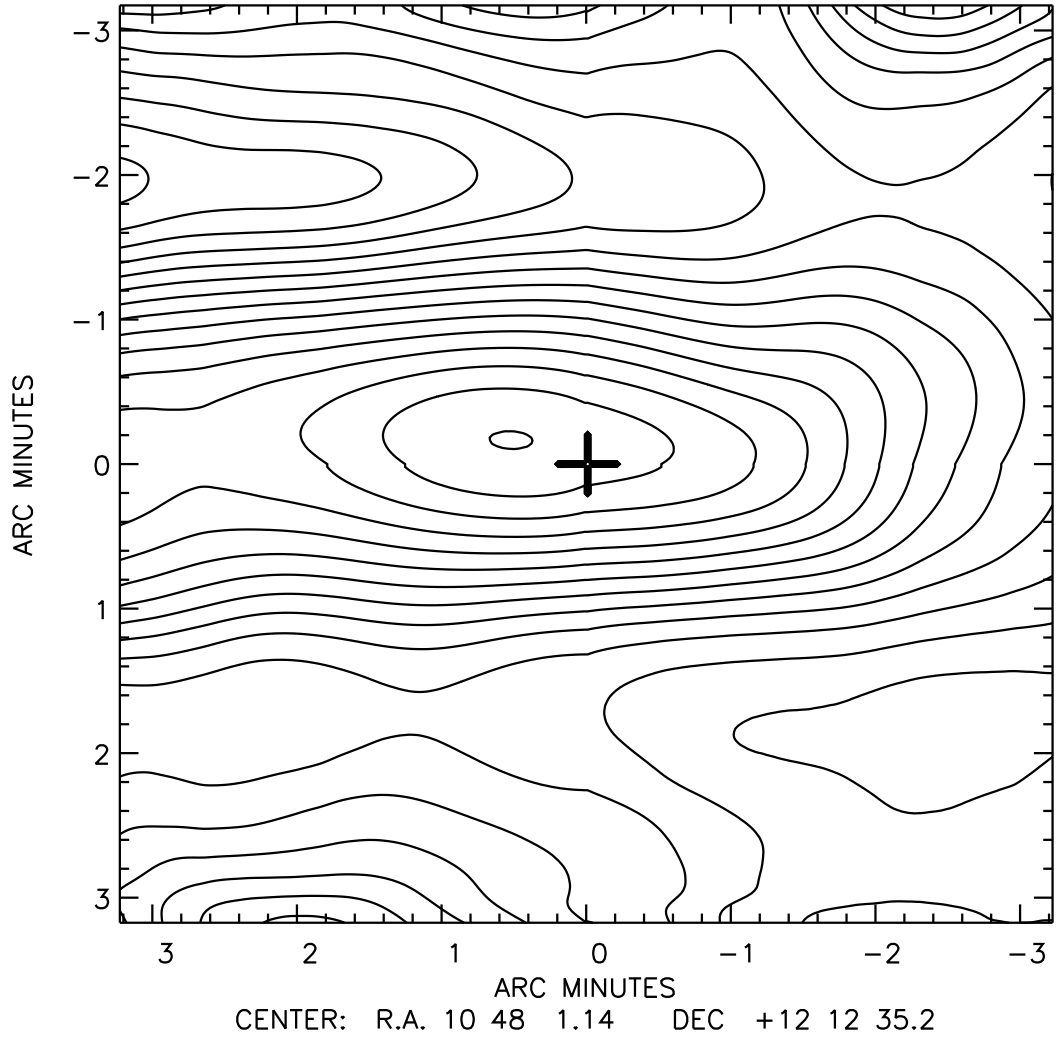}
\caption{Result from the cross correlation analysis between the 8$\mu$m and the H{\sc i} maps in the field of the highest density structure in the Leo cloud. The central position is noted by a cross and corresponds to where the maximum of correlation would be expected if the spatial distributions were similar. 
\label{fig5}}
\end{figure}

We performed a normalized cross correlation analysis between the 8$\mu$m map and the HI map in the field where the high density peak is observed\footnote{The normalized cross correlation of an image $f(x,y)$ by the HI template $t(x,y)$ is given by $\frac{1}{n-1}\Sigma \frac{(f(x,y)-\bar{f}))(t(x,y)-\bar{t})}{\sigma_f\sigma_t} $ where $\sigma$ is the standard deviation of the signal and the sum is done on the $n$ pixels $(x,y)$ of the overlapping region between the map and the template.}. The result of the cross correlation is shown in Fig. \ref{fig5}. If the 8$\mu$m emission in this region is dominated by dust emission associated with the Leo cloud, then one would expect the spatial variations in both maps to be similar and therefore the cross correlation to be maximum at position (0,0) (represented by a cross in the center of Fig. \ref{fig5}). We observe that the maximum of spatial correlation is shifted with respect to what is expected by $\sim40"$,  which is just slightly less than the VLA beam size. Note, however, that the 8 micron emission peak is well-centered within the next lower HI column density contour, which is more suggestive of an association.

While the offset in Fig \ref{fig5} might be the result of the astrometric and flux level uncertainties in the H{\sc i} data due to the digitalization of the contour image, one cannot rule out the possibility of the Spitzer detection being unrelated to the Cloud. The most likely scenarios would be the fortuitous presence of a cirrus clump or an overdensity of extragalactic sources along the line of sight to the Cloud. The first scenario could be tested with a deep survey for H{\sc i} emission from the Milky Way, and the second might me tested with deep visible and near-infrared imaging to look for the extragalactic sources. Since neither data set is available at this time, the association of the detection with the Cloud needs to remain a tentative result.

\subsection{Comparison with GALEX data}

The recent detection of possible star formation sites in the Leo Cloud \citep{Thilker:2009kx} in the UV is important with respect to our possible detection of dust emission. Indeed, one of the reported region (clump 1) is located inside the highest HI density contour where we also detect dust emission, and could play a role in the dust heating.

However, the peak of dust emission is displaced by $\sim 1.5'$ ($\sim 4$ kpc at the distance of the M96 group) with respect to the reported star forming region in clump one. Furthermore, the UV emission is extremely diffuse (the estimated stellar density is $4\times 10^7M\odot$ in 4.1(kpc)$^2$) and the estimated flux density in this region is comparable to the one of the surrounding galaxies as seen from the cloud (the shape and intensity of the radiation field produced by the star forming region will be discussed in sect. \ref{sec:discussion}).

\section{Discussion\label{sec:discussion}}

Even though the dust emission is barely detected, it is potentially of great significance in providing clues to the origin of this extragalactic gas cloud. How much dust might be inferred to be held in the Leo Cloud?

\begin{figure}
\epsscale{.80}
\plotone{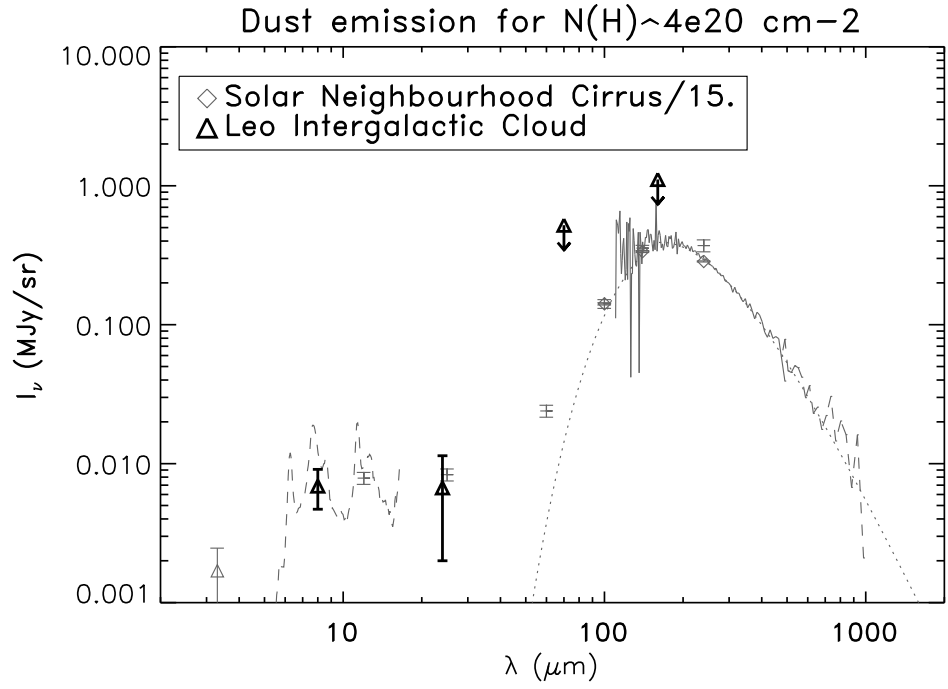}
\caption{Spectral energy distribution of dust emission associated with the densest H{\sc i} in the Leo intergalactic cloud (black triangles and upper limits). It is compared to the spectral energy distribution observed in the solar neighborhood (curves and points in grey), scaled to a H{\sc i} column density of $4\times 10^{20} at.cm^{-2}$ and divided by a factor of 15.\label{fig6}}
\end{figure}

We build a spectral energy distribution of the infrared emission from the brightest H{\sc i} bin of the Leo Cloud, including both detections and upper limits. This spectral energy distribution is compared in Fig. \ref{fig6} to the one observed in the Solar Neighborhood, normalized to $N_{HI}=4\times 10^{20} \mathrm{cm}^{-2}$. The local emission is a combination of results from \citet{BRA+96,GLP+94,BBD+94} and \citet{BAB+96}. Because we are concerned primarily with 8 and 24 $\mu$m emission, which is dominated by fluctuating grains, we may assume that the emission is proportional to the product of the total dust column with the heating intensity, or equivalently the total H{\sc i} column density times the dust to gas ratio times the heating intensity.

As Fig \ref{fig6} shows, dust emission potentially associated with the Leo Cloud is about 15 times weaker than the local emission for the same H{\sc i} column density, but has infrared colors consistent with the local emission. If the dust in the Leo Cloud were heated by an interstellar radiation field similar to the one in the Solar Neighborhood, one would infer that the dust to gas ratio in the Leo Cloud is 15 times lower than in the galaxy.

\begin{figure}
\epsscale{.80}
\plotone{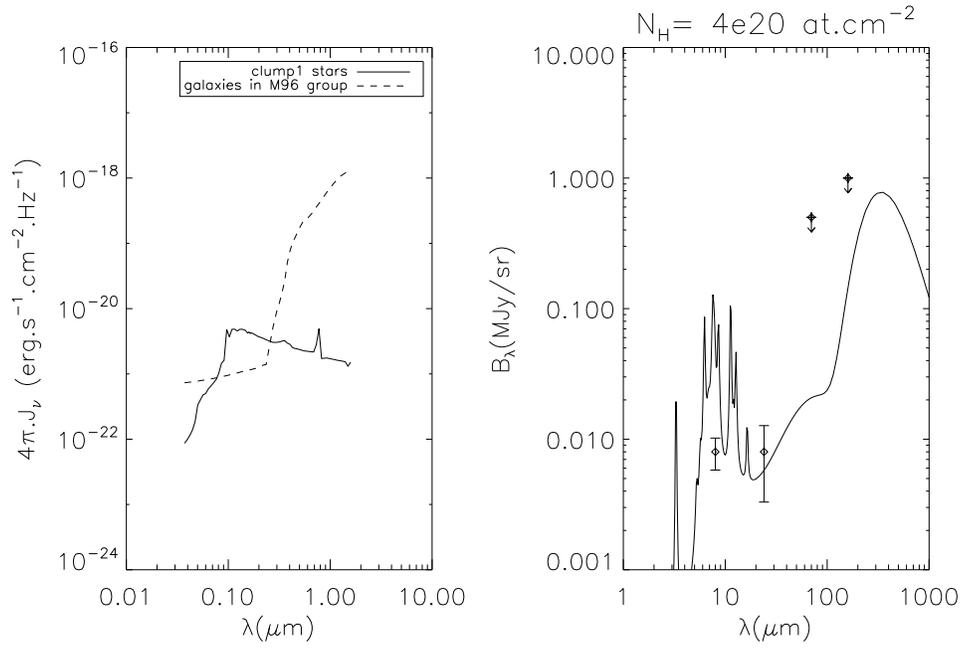}
\caption{Spectral energy distributions of the radiation fields heating the clump where dust is detected (left panel) and resulting estimated dust emission (right panel), scaled by a factor of 5 and to an H{\sc i} column density of $4\times 10^{20} at.cm^{-2}$ to fit the observed infrared emission (diamonds).\label{fig8}}
\end{figure}

However, there are two other factors that enter this estimation, with opposite sign implications for the dust to gas ratio:
\begin{itemize}
\item First, the Leo Cloud is illuminated by stellar light from the neighboring galaxies and potentially also from the newly detected star forming region in the cloud. \citet{Schneider:1989rt} estimated the heating contributed by the surrounding galaxies and found it to be about 24 times lower than the heating in the Solar Neighborhood. The main uncertainty on this value is the shape of the radiation field from the galaxies. \citet{Schneider:1989rt} assumed a spectral energy distribution similar to the one in the solar neigborhood \citep{MMP83}. Since the 8 and 24$\mu$m emission from transiently heated particles is sensitive to the hardness of the radiation field, we checked the importance of this effect. The radiation field spectral energy distribution due to the three major galaxies surrounding the Leo Cloud was estimated using fluxes from the UV to near-IR collected with the NASA/IPAC Extragalactic Database (NED). Since the M96 group is at high galactic latitude, the extinction from the dust in our galaxy is negligible and we did not correct for that effect. The UV/optical ratio obtained is lower than the one observed in the solar neighborhood, as expected from the large extinction from dust inside the galaxies. To estimate the intensity and shape of the radiation field coming from the star forming region in clump 1, we used a synthetic spectrum computed with the GALEV model \citep{Kotulla:2009ve} using the best model parameters obtained by \citet{Thilker:2009kx}. Without more knowledge on the distance between the UV sources and the dust cloud, we assumed that both emission are coincident and that the radiation field is heating the dust isotropically. Both estimated radiation field are presented in Fig. \ref{fig8},left panel. While they have very different UV to optical shapes, the energy from the UV emitting region is comparable to the one of the surrounding galaxies as seen from the cloud.  We used the  \citet{DBP90} dust model with the computed radiation field to estimate the resulting infrared emission and in such a case, we would need as much as 5 times the solar dust abundances with respect to the gas to explain the 8 and 24$\mu$m emission in the Leo cloud (Fig. \ref{fig8}, right panel). However, this estimate of the radiation field is highly depending on the assumptions on distance and geometry and the relative contribution betwen the star forming region and the nearby galaxies which will change the UV/optical ratio. We estimated that the change in the radiation field hardness could affect the 8 and 24$\mu$m  dust emission by a factor 1.5 and 4 respectively.).  
\item Second, the Cloud is likely to have such low density that collisional excitation of the H{\sc i} falls short of thermalizing the 21cm transition levels and the detected H{\sc i} signal underestimates the true total column density  \citep{Schneider:1983yw}. Indeed, the apparent stability of the Cloud against disruption argues for a total mass about 13 times greater than the mass derived from the H{\sc i} line brightness. On the other hand, \citet{deguchi:1985gd} show that trapping of Lyman-$\alpha$ photons generated by extragalactic ionizing radiation can prevent the suppression of spin temperature if the column density is high enough and the velocity dispersion is low enough. Given the column density and velocity spread within the clump that may possibly have been detected, the spin temperature would likely be thermalized.
\end{itemize}
Our conclusion is that the dust to gas ratio in the Leo Cloud could be as high as a few times the solar value, but it is unlikely to be smaller than 1/20 solar if the detection is indeed of the Cloud. If confirmed, this would render unlikely a primordial origin for the Cloud. 

\section{Conclusion}

Our analysis of infrared emission in the direction of the Leo intergalactic cloud shows a possible detection of dust emission at 8$\mu$m and a marginal detection at 24 $\mu$m associated with the brightest H{\sc i} clump of the cloud. At 70 and 160$\mu$m, upper limits on the dust emission were deduced.

Variations in the infrared emission not associated with the Leo Cloud are large and one can not rule out the possibility of a spurious detection without further observations. However, if this detection is confirmed, the quantities of dust inferred imply a dust to gas ratio in the intergalactic cloud no less than 1/20 the solar neighborhood value and up to a few times solar. This result is important as it would exclude the possibility that the intergalactic cloud has a primordial origin. Instead, this large intergalactic cloud could therefore have been formed through collision between galaxies in the group or through the stripping of a low surface brightness galaxy by the group potential as suggested by \citet{Bekki:2005eu}.

\acknowledgments

This work is based on observations made with the Spitzer Space Telescope, which is operated by the Jet Propulsion Laboratory, California Institute of Technology under a contract with NASA. This research has made use of the NASA/IPAC Extragalactic Database (NED) which is operated by the Jet Propulsion Laboratory, California Institute of Technology, under contract with the National Aeronautics and Space Administration. We would like to thank M. Fouesneau and D. Thilker for their help on the estimate of the UV radiation field from the star forming region with evolutionnary synthesis models.

{\it Facilities:} \facility{Spitzer}.

\bibliographystyle{aa}
\bibliography{../../biblio}

\newpage

\cleardoublepage
\newpage

\end{document}